# The Kinetics of Dissolution of an Amorphous Solid

Ian Douglass and Peter Harrowell[*]

School of Chemistry, University of Sydney, Sydney, NSW 2006 Australia

*peter.harrowell@sydney.edu.au

Abstract

The kinetics of dissolution of an amorphous solid is studied using a simple model of a glass that captures with reasonable accuracy the dynamic heterogeneities associated with the relaxation of an amorphous material at low temperatures. The intrinsic dissolution rate is shown to be proportional to the concentration of surface particles kinetically able to exchange with the solvent, independent of temperature or the thermal history of the glass. The morphology of the dissolving surface is described and the possibility of using surface etching to image dynamic heterogeneities is explored.

## 1. Introduction

The dissolution kinetics of amorphous materials is fundamental to a number of important processes. A significant number of pharmaceuticals must be delivered in the amorphous state in order to have dissolution rates fast enough to be effective[1-3]. The selective dissolution of one metal species from the surface of an amorphous alloy (i.e. a metallic glass), either through the oxidation of one species or by selective solvation, is being used to fabricate nanoporous electrodes and catalysts[4,5]. Selective dissolution is central to understanding the dissolution of silicate-based glasses[6-9]. In these materials, the transition from solid to solute typically proceeds via an intermediate gel state. The dissolution of amorphous polymers can also exhibit an analogous scenario in which the first step of the processes involves the penetration by the solvent at rate faster than the relaxation of the entangled polymer network. This gives rise to what has been classified as *Case II* diffusion[10,11]. (Case I being regular or Fickian diffusion where the relaxation rate is faster than the solvent diffusion.)

As dissolution is a phenomenon associated with a solid, a fundamental treatment of the dissolution of an amorphous solid must address the nature of solidity of glasses. An amorphous material can exhibit mechanical response that varies continuously with temperature between that of a liquid and a solid. What, then, might we expect of the kinetics of mixing when an amorphous material is put in contact with a liquid solvent? While two liquids mix via *inter-diffusion*, a solid (amorphous or crystalline) must undergo some type of *dissolution*. As the distinction between the two modes of mixing arises entirely from the kinetics of the glass forming species, it is necessary to treat this kinetics with a physically realistic model. Starting with work in 1991 [12,13] based on a version of the model presented below, it is now widely appreciated that the dynamics of amorphous materials is characterised by striking transient heterogeneities arising from the complex coupling between spatial fluctuations in structure and the dynamics[14]. The goal of this paper is to present a model of the dissolution of an amorphous material in which the process of dissolution and solvent mixing is treated on the same footing as the kinetics of structural relaxation in the amorphous material and the associated dynamic heterogeneities..

This paper is organised as follows. In the next Section we introduce a number of basic concepts regarding amorphous dissolution including the notion of the intrinsic dissolution rate. In Section 3 we present our model of the amorphous state and the interaction between solvent and glass forming solute. The temperature dependence of the intrinsic dissolution rate is reported in Section 4, along with an analysis of the role played by the degree of stability (the so-called fictive temperature) of the amorphous configurations. In Section 5 we examine the surface morphology associated with amorphous dissolution and the relationship between this structure and the dynamic heterogeneities of the amorphous solid. The cross over in dynamics between dissolution of the rigid amorphous state and the inter-diffusion of the fluid amorphous state is discussed in the Appendix.

**2. The Phenomenology of Amorphous Dissolution**

The dissolution of a solid can be characterised by the kinetics, as measured by the time dependence of the solute concentration in solution , *dc/dt*, and by thermodynamics, as measured by the equilibrium concentration $c_{eq}$ at saturation. The two aspects of dissolution are referred to as *dissolution rate* and *solubility*, respectively. In Fig. 1 we depict, qualitatively, both features of dissolution and compare the dissolution of an amorphous solid and its corresponding crystal. Note that the dissolution rate decreases in time, to vanish at saturation. It is useful, therefore, to characterise the *intrinsic* dissolution rate as the rate of dissolution at c = 0 (see Fig. 1), i.e. the dissolution rate into the pure solvent. The solubility (i.e. the asymptotic concentrations in Fig. 1) is generally larger for the amorphous solid than the crystal due to the larger enthalpy of the former; just how larger depends on the enthalpy difference between the two types of solid and the degree of solvent inclusion in the two solids. The depiction in Fig. 1 is something of an idealization and the degree to which it can be experimentally realised is discussed below in conjunction with Fig. 2.

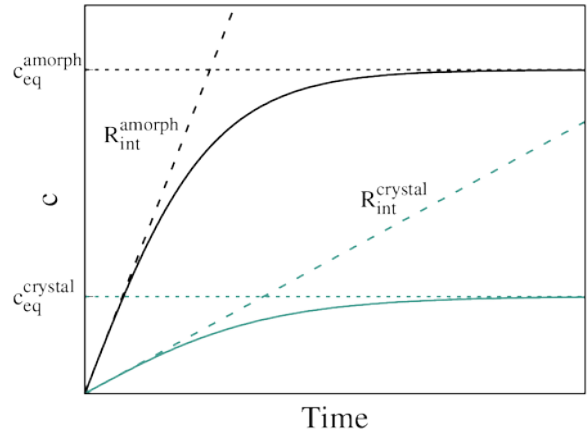

**Figure 1.** Sketch of the dissolution curves for an amorphous (black) and crystalline solid (teal). The respective $c_{eq}$'s are indicated by horizontal dotted and the intrinsic dissolution rates correspond to the slope of the curve c(t) at c = 0 (black and teal dashed lines).

We can write the dissolution rate $R(t) = dc(t)/dt$ as

$$R(t) = k \cdot A(t) \cdot [1 - \exp(\beta \Delta \mu(c))] \tag{1}$$

where $\beta = (k_B T)^{-1}$, A is the microscopic surface area (a quantity that can change with time) and $\Delta \mu(c) = \mu_{sol}(c) - \mu_{solid}$, where $\mu_{sol}(c)$ is the chemical potential of a solute particle in solution at a concentration c and $\mu_{solid}$ is the chemical potential of the solute in its solid form. Note that $\Delta \mu(c_{equ}) = 0$, by definition. The kinetic coefficient *k* is the rate per unit area at which solute particles leave the solid for the solution.

The concentration dependence of the rate, already noted in Fig. 1, arises in Eq. 1 through the dependence of $\Delta \mu(c)$ on c. In general, the dependence is complicated by the spatial gradient in the solute concentration in solution associated with the diffusive transport of solute from the solid surface into the solution. As we are interested in how the properties of the amorphous solid influence the dissolution rate, we shall avoid the problem of the time evolution of the solute concentration by focusing on the intrinsic dissolution rate $R_{int}$, defined as the rate of dissolution at c = 0, i.e.

$$R_{int}(t) = k \cdot A(t) \cdot [1 - \exp(\beta \Delta \mu(0))] \qquad (2)$$

The introduction of the intrinsic dissolution rate here closely parallels the approach used to model crystal dissolution. This is the case, for example, in a popular model of crystal dissolution, the kinetic Monte Carlo model based on the Ising spin lattice[15]. Here dissolution of a particle is modelled as a transformation from solute to solvent (i.e. a 'spin flip'). Adsorption is allowed in this model via the reverse processes but the solute concentration remains zero, by construction, at all times. Experimentally, a dissolving solid interface can be maintained in a state close to the c = 0 limit by applying a solvent flow to convect the solute away from the surface[16-18]. The intrinsic dissolution rate would represent the large flow rate limit. (It is assumed here that the solvent flow rates are always well below those required to mechanical erode the interface[19,20].) Exactly how the c = 0 condition is imposed in a simulation of dissolution involves some subtleties in terms of defining exactly when a solute particle has 'detached' from the solid. We shall return to this point in Section 4.

While we shall only consider the amorphous dissolution in this paper, the difference between the amorphous and crystalline solids is of central interest to the applications in pharmaceuticals and so we shall conclude this Section with a brief discussion of this issue. The difference in solubility between the amorphous and crystalline solids (as measured by $c_{eq}$) arises from the lower free energy of the crystal relative to the glass. Hancock and Parks[21] expressed this relation as

$$\frac{c_{eq}^{amorph}}{c_{eq}^{crystal}} = \exp(\Delta \mu^{am-cryst} / k_B T) \qquad (3)$$

where $c_{eq}^{amorph}$ and $c_{eq}^{crystal}$ are the concentrations of solute at coexistence with the amorphous and crystal solids, respectively and $\Delta\mu^{am-cryst} = \mu^{amorph} - \mu^{crystal}$, the difference in chemical potentials of the amorphous and crystal solids. In the context of drug delivery, the rate of dissolution is particularly significant since, in many cases, it determines the therapeutic value of a compound[1-3]. As shown in Fig. 1, the rate of dissolution, dc/dt, is strongly influenced by the thermodynamic driving as measured by $c_{eq}$. This means that the benefit of an amorphous drug preparation in terms of faster dissolution arises largely from the instability of the glass relative to the crystal and glass as measured by $\Delta\mu^{am-cryst}$.

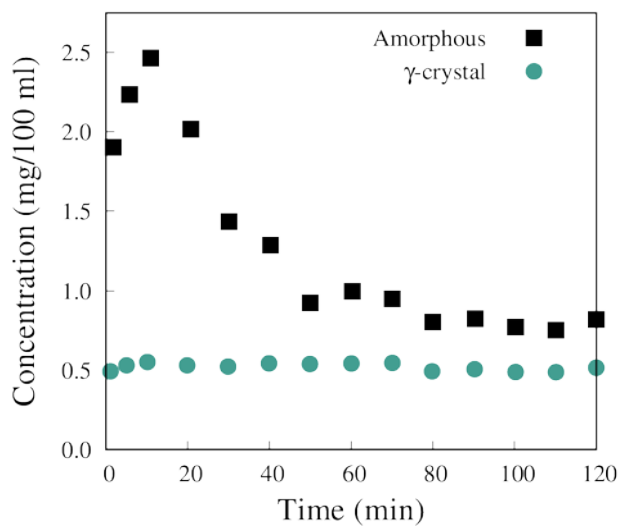

**Figure 2.** Experimental dissolution curves for amorphous and crystalline indomethacin at 25°C from ref. 21. The curves represent the concentration of solute in solution as a function of time following the immersion of the solid (crystal or amorphous) into the solvent. A constant concentration indicates saturation has been reached.

There is, however, a problem associated with the idealized dissolution curve for the amorphous solid in Fig. 1. Since the amorphous solid is not an equilibrium state, the very notion of $c_{eq}^{amorph}$, i.e. a quantity associated with the equilibration of the solution with the amorphous solid, is questionable. $c_{eq}^{amorph}$ is typically not directly observable because, as the glass dissolves, the solute simply precipitates back out of solution as the crystal. An example of this behaviour is provided by the measurements of the dissolution of indomethacin[21] as shown in Fig. 2. Polymer additives to the solution can supress the crystal growth[22] and so permit the 'true' solubility advantage of the amorphous solid (i.e. as given by Eq. 1) to be observed. In terms of modelling of amorphous dissolution, it is clear from this discussion that we would like to consider a model which does not have a competing equilibrium crystal phase. Such a model is presented in the following Section.

## 3. Model

The goal of this work is to treat the dissolution kinetics consistently with the structural relaxation dynamics of the amorphous material. To this end our model is based on a simple lattice model of glass-like relaxation introduced by Fredrickson and Andersen[23,24] in 1984. In this model, known as the facilitated kinetic Ising (fkI) model, a volume is discretized into the sites on a periodic lattice. We shall consider a simple cubic lattice in 3D. A numerical study of the relaxation kinetics of this 3D fkI model has been reported by Graham et al[25]. The 'kinetic influence' of the material at each site is characterised by a property that can have just one of two values. It is convention to call this property a 'spin' and the two values correspond to the spin being 'up' ($\sigma = 1$) or 'down' ($\sigma = 0$). The concentration of up spins is determined by the equilibrium condition

$$c = \exp(-\beta h)/(1 + \exp(-\beta h)) \qquad (4)$$

where h is the energy increase associated with flipping a down spin up. In the following, we shall use reduced temperature units such that $h/k_B = 1$. The role of these spins is to establish the link between a configuration and the relaxation dynamics. The probability that the spin on site i will flip is

$$W_i(\sigma_i, m_i) = \exp[(\sigma_i - 1)/T] H(m_i - 3) \tag{5}$$

where $\sigma_i$ refers to the current spin state in site i, $m_i$ = number of neighbours with up spins and the Heaviside step function, $H(x) = 0$ for $x < 0$ and $= 1$ for $x \geq 0$. The transition probability requires that a site has three or more up neighbours in order to flip and that down spins are increasingly favoured as T decreases.

The fkI model provides a generic representation of the consequences of dynamics determined by the local structure. The individual sites represent individual molecules – solute or solvent – assumed to be uniformally distributed in space (i.e. we do not consider density fluctuations in this model). The spin represents a local measure of the configurational packing about each particle with up and down spins corresponding to poorly packed and well packed, respectively. The influence of the packing efficiency is manifest by its influence on the kinetics. Here, we extend this model to include the solvent as follows. Our lattice sites can be occupied by one of two types of particles: the original glass former ('p' particles) or a solvent particle ('s' particles). As in the original model, the p particle has a spin $\sigma = 1$ or 0 while the s particle has a spin $\sigma = 1$ only. By this means, solvent particles always enhance the likelihood a spin flips in adjacent p particles. The probability of spin flips on p particles still obeys the transition probability in Eq. 5. An s particle on lattice site i and a p particle on neighbouring lattice site j can swap places with the following probability:

$$T_{ij} = H(m_i - 3)H(m_j + \sigma_j - 4)\min\{1, \exp(-\Delta E_{ij}/T)\} \tag{6}$$

$\Delta E_{ij}$ is the change in energy associated with the s-p swap and it equals the change in the number of s-p 'bonds' $\Delta n_{ij}$ = number of s-p neighbour pairs associated with sites i and j *after*

swap - number of s-p neighbour pairs associated with sites i and j *before* swap. The energy change is

$$\Delta E_{ij} = J \times \Delta n_{ij} \tag{7}$$

If $J < 0$ then we are favouring forming s-p bonds and hence the mixing of the s with the p is energetically favoured (an exothermic process). If $J > 0$, then s-p interactions are discouraged and the two types of particles will tend to demix. In this study we shall set $J = 0$ so as to concentrate on the T dependence arising from the kinetic of relaxation. (Elsewhere we shall consider $J > 0$ and the precipitation of our glass forming solute.) Note that a temperature dependence remains when $J = 0$ due to the energy *h* associated with configurational changes in the glass forming solute.

The particle-solvent model presented here is similar to previous extensions of the fkI model considered by Schulz and coworkers[26,27]. In ref. 26, the authors consider two components whose motion is determined by the spin field. In ref. 27, one of the components is regarded as a vacancy with a neutral contribution to the spin field. The treatment we present here differs from either of these previous versions in that our solvent species contributes positively to the facilitated motion (i.e. the solvent is regarded as a 'plasticizer') and we are specifically interested in the evolution of an initial step function in the concentration. The initial interface between solid solute and solvent is sketched in Fig. 3. The picture is similar to that used previously to model the glass-vacuum interface[28], the difference being that in mapping the vacuum sites to solvent particles we have added the interaction strength as given in Eq. 7.

The algorithm now goes as follows: 1) Randomly select a site. 2) If the site is occupied by a p particle then a spin flip is attempted as in the previous algorithm. 3) If the site is occupied by an s particle then randomly select a neighbour site j; if the site j is occupied by another s particle do nothing and if site j is occupied by a p particle calculate the change in s-p pairs if the swap took place. 4) Accept the swap with a probability $T_{ij}$, and then randomly select the next site, and so on. Time is measured in terms of Monte Carlo cycles where one cycle is equal to M trial moves where M is the number of lattice sites.

The solvent-solute exchange can only take place between two sites that both satisfy the spin flip condition. Dissolution, in other words, is bound by the same cooperativity that constrains relaxation in the glass-former itself. Microscopic reversibility is ensured by Eq. 6 so that any move that would result in a particle finding itself in a site that cannot undergo the reverse process is rejected. We shall use a simple cubic lattice in 3D of size 75x75x60, unless otherwise specified. Previous studies[25] of this system have determined the glass transition temperature $T_g = 0.46$. Periodic boundaries were used in all three directions. The z coordinate was initially split into two sections – one of pure solvent, and one of pure glass.

An implicit feature of our model is that the solvation of a particle p will generally require the particle be in the spin up state. To see why, consider a cube of p particles, all spin down, surrounded by solvent. The p particles on the corners of the cube have 3 solvent neighbours and so are flippable. If we were to exchange one of these corner particles with a neighbouring solvent, then the solvent would find itself with 4 down neighbours (the 4th down neighbour being the p particle that has been exchanged), an unflippable environment and, hence, forbidden by microscopic reversibility. It could be argued that the spin state of the p particle should not matter once it is solvated but any rule that ignored the spin state of the dissolving solute or allowed it to flip during the exchange would run the risk of violating reversibility.

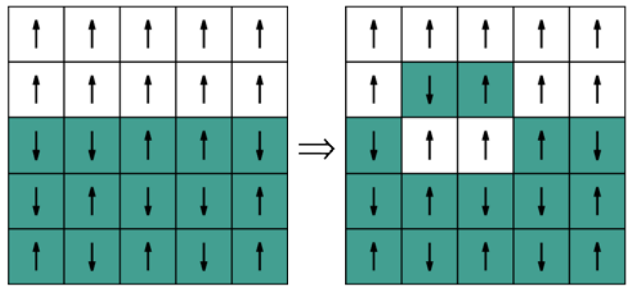

**Figure 3**. An illustration of solvent penetration into the glass phase. Solvent particles are shown in white, and are always up spin. Solute particles are in teal, and can have up or down spin. Only solute particles with a requisite number of up spin neighbours can be swapped.

## 4. Intrinsic Dissolution Rate

The rate at which the concentration in solution of a solute increases through the dissolution of the solid solute reflects the combination of two distinct dynamic processes: the intrinsic dissolution rate at which solute particles are extracted from the solid surface, and the diffusive dispersion of solute through the solution. The latter process, in which a moving interface is coupled to a diffusive field is referred to as the Stefan problem[29-31] and is common to all dissolution processes, independent of whether the solid is amorphous or crystalline. For this reason we shall focus on the intrinsic dissolution rate in this paper and, in this Section, we shall examine the factors that influence this rate

To separate out the intrinsic dissolution we must 'turn off' the deposition of solute particles from solution. To do this we shall constrain the solution concentration to be zero. This is achieved in our model by converting each solute particle into a solvent once it is deemed to be detached from the solid surface. We shall define a solute particle detached once it has 6 solvent neighbours. While this algorithm allows us to define *an* intrinsic dissolution rate, it is not clear how it should be coupled to the extended diffusion field in the solution since such a coupling would influence concentration within the interface region. We shall consider the question of this coupling at the end of this Section.

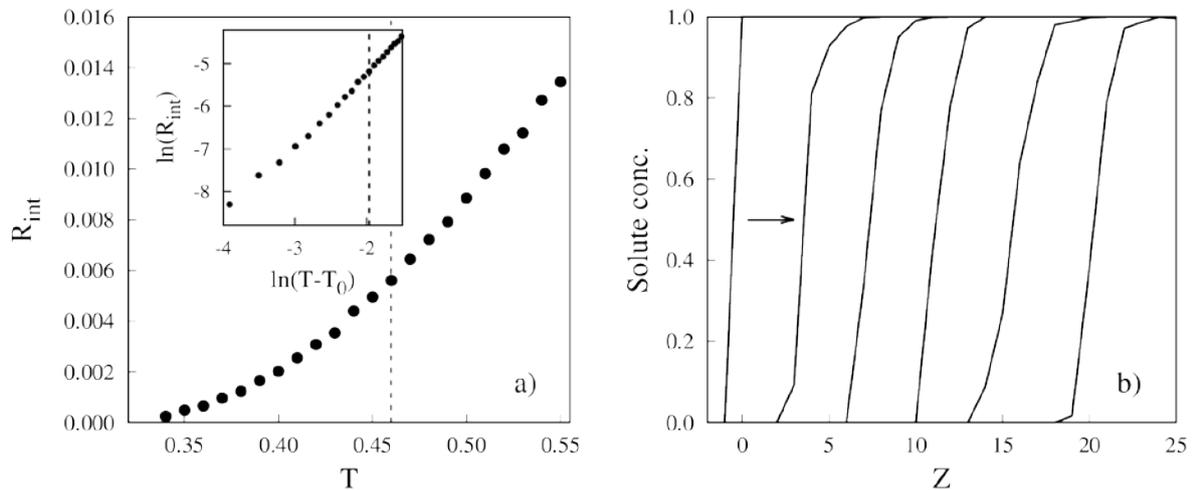

**Figure 4.** a) Intrinsic dissolution rate of the *equilibrium* amorphous solid as a function of T with the bulk glass transition temperature $T_g$ indicated by the vertical line. Insert: Plot of $\ln(R_{int})$ vs $\ln(T-T_o)$ where $T_o \sim 0.32$ is the temperature at which $R_{int}$ appears to vanish. b) A time sequence of interface profiles (evolving from left to right as indicated by arrow) showing the front-like dissolution below $T_g$.

The temperature dependence of the intrinsic steady state dissolution rate per unit area $R_{int}(t)$ of the equilibrium amorphous solid is plotted in Fig. 4. While we focus on $T < T_g$, we have included data for temperatures above $T_g$ just to establish the continuity of $R_{int}$ to higher temperatures. The transition from glass to liquid on heating above $T_g$ must have its analogue in the cross-over of the mixing process from dissolution to inter-diffusion. In the Appendix, we examine this general feature of the problem. With regards Fig. 4, we note the following features. i) Below $T_g$, dissolution proceeds with a concentration profile that propagates in a front-like fashion as shown in Fig. 4b. ii) Dissolution becomes too slow to observe for T < $T_o$ where $T_o \sim 0.32$ and increases with T as

$$R_{int}(T) \propto (T - T_o)^\gamma \tag{8}$$

with the exponent $\gamma = 1.69$ extracted from the fit of Fig. 4 (see upper insert).

What is the origin of this critical temperature $T_o$? As we have set the enthalpy of mixing to zero (i.e. by setting J = 0 in Eq.7), all temperature dependence of the dissolution rate must arise from the energy associated with configuration fluctuations of the solute particles (represented here by the solute spin states). The condition for the exchange of solute and solvent particles requires that the solute particle is mobile in terms of the facilitated kinetics of the glassy material. We expect, therefore, that the dissolution kinetics should depend strongly on the concentration $c_m^s$ of mobile solute particles at the surface of the amorphous solid. In the context of our model, solute mobility does not exactly correspond to ability of that solute to be exchanged with a solvent. To understand this, consider the cases sketched in Fig. 5.

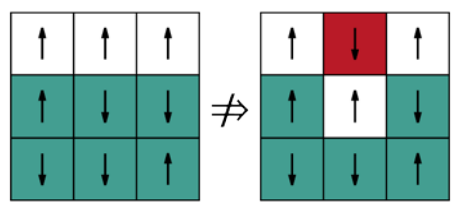

**Figure 5.** An example of a solute particle (indicated in red) that is mobile (as defined by the facilitation condition) but that cannot exchange with a solvent as it would result in an immobile solvent and hence violate microscopic reversibility.

At low temperatures, surface particles that can be exchanged are those that satisfy *both* the mobility condition and either have an up spin or have at least 3 solute neighbour with up spins. We shall call the concentration of these surface exchangeable particles $c_{ex}^s (< c_m^s)$. While these details of the model, e.g. the retention of the configurational variable represented by the spin on a solute that has become solvated, may not have any direct correspondence with the actual physical situation, the general idea that mobility associated with structural relaxation and the capacity of a surface particle to exchange with the solvent can be related but different seems reasonable.

We have calculated $c_{ex}^s$ for the dissolving surface at steady state. In Fig. 6a we plot the dissolution rate against $c_{ex}^s$ and find that the dissolution rate is proportional to $c_{ex}^s$ at all temperatures, confirming that the apparent cessation of dissolution at low T in Fig. 4 is a direct consequence of the disappearance of exchangeable solutes at the surface of our glass.

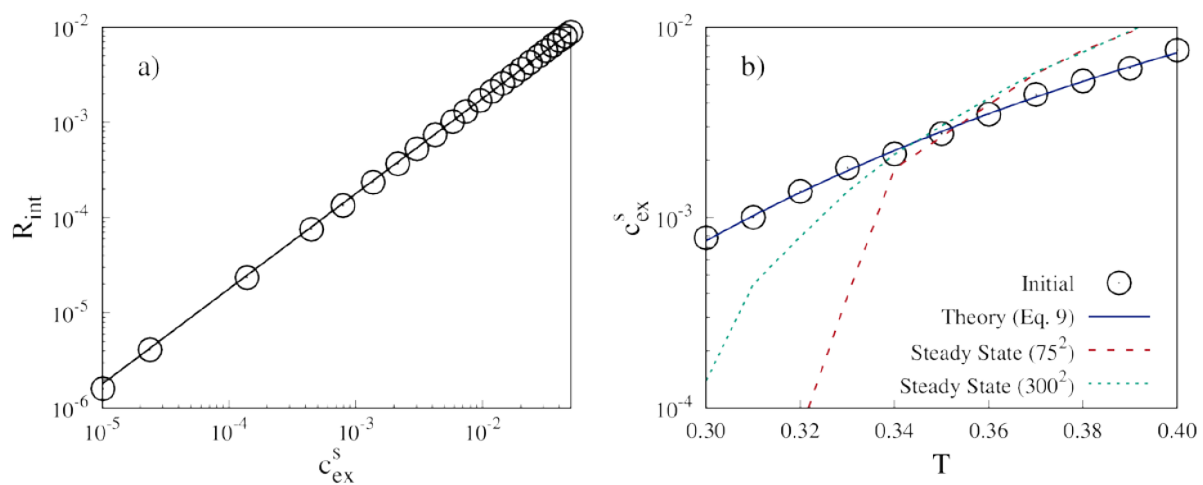

**Figure 6.** a) Plot of dissolution rate against $c_{ex}^s$ for the equilibrium amorphous solid. The solid line is a fit of the form $R_{int} = 0.18 c_{ex}^s$. b) $c_{ex}^s$ vs T for the initial surface (red circles) and the steady state surfaces with dimensions 75x75 (red dashed curve) and 300x300 (teal dotted

curve). The theoretical result from Eq. 9 is plotted as a blue line and agrees closely with calculated values of $c_{ex}^3$ of the initial surface.

In Fig. 6b we plot the temperature dependence of $c_{ex}^s$ for both the initial surface and the steady state dissolving surface. We find that $c_{ex}^s$ of the initial surface is well described by the theoretical equilibrium expression for $c_{ex}^s$ calculated assuming a smooth surface solid surface, i.e.

$$c_{ex}^s = 10c^3[(1-c)^2 + (1-c)^3] + 5c^4(1-c) + c^5 \qquad (9)$$

As shown in Fig. 6b, Eq.9 accurately reproduces $c_{ex}^s$ for the smooth glass surface but not the steady state surface. At high temperatures, the steady state concentration of exchangeable solute particles is higher than that predicted for a smooth surface due to the roughening associated with dissolution which, by effectively increasing the surface area, increases the number of mobile solute particles. At low temperatures, we observe a crossover in which the steady state value of $c_{ex}^s$ below that of the equilibrium smooth surface, to vanish at the $T_o$ observed in Fig. 4. The equilibrium expression, in contrast, predicts that $c_{ex}^s$ only vanishes at T = 0. To understand what happens here it is useful to appreciate that with each exchange of solute and solvent at the interface, a) a mobile solute particle is removed, and b) the newly inserted solvent particle acts to mobilise additional solute particles. For dissolution to propagate, the concentration of mobile solute particles must be sufficient to span the surface. This is related to the problem of the loss of ergodicity in the facilitated kinetic Ising model at sufficiently low concentration in a system of finite size[23,24]. If the concentration of extractable solutes drops below this threshold concentration, dissolution is not possible and all that we observe is the removal of those extractable solute particle by solvent exchange.

This is a finite size artefact of the model for cooperative dynamics and is associated with the physical inaccessibility of the low temperature configurations we are using in these calculations. A more realistic protocol is to consider the case where the fictive temperature $T_f$ of the amorphous solute is chosen to be around $T_g$, corresponding to mobile solute concentrations well above this ergodocity breaking threshold. We discuss the dissolution of these out of equilibrium glasses in the following. We shall assume that the frozen-in glass configuration corresponds to the equilibrium configuration at $T_g$. The effective temperature of an out of equilibrium configuration is referred to as the *fictive* temperature, $T_f$ [32,34] and is defined here in terms of the spin up concentration by $T_f = 1/\ln(c^{-1} - 1)$, i.e the inverse of the equilibrium relation in Eq. 4. It is possible to produce amorphous solids with a range of fictive temperatures. A high fictive temperature can be generated by very rapid quenching or mechanical grinding of a sample[35], while sub-$T_g$ samples can be generated by aging[36] or by vapor deposition[37] of the amorphous material.

We can model the effect of a non-equilibrium glass by selecting configurations associated with a fictive temperature other than that of the temperature itself. In the following we shall consider fictive temperatures $T_f = 0.9T_g$, $T_g$ and $1.1T_g$, as this range covers the likely values of $T_f$ that might be physically realised. (A larger range of $T_f$ is possible if more extreme measures are taken: higher $T_f$'s by extremely rapid quenching and lower $T_f$'s by vapor deposition or ultra-slow cooling, for example.) Our first question is whether the density of exchangeable sites still determines the intrinsic dissolution rate, irrespective of their deviation for equilibrium. As shown in Fig. 7a, we find exactly the same linear relation between $c_{ex}^s$ and $R_{int}$ as we found previously in Fig. 6a. What has changed with the choice of $T_f$ is the relationship between $c_{ex}^s$ and T. As is shown in Fig. 6b, at $T < T_f$, the high energy configurations result in higher values of $c_{ex}^s$ and, hence, higher dissolution rates. This results in the critical temperature $T_o$ at which dissolution ceases to be significantly decreased, relative to the equilibrium case. The magnitude of the influence of $T_f$ can be gauged by looking at the ratios $R_{int}(T,T_f=1.1T_g)/R_{init}(T,T_f=T_g)$ and $R_{int}(T,T_f=0.9T_g)/R_{init}(T,T_f=T_g)$ which we find to equal 1.4 and 0.6, respectively, independent of T. These ratios represent our estimate of the range of variation one might expect in dissolution rates for different glass histories.

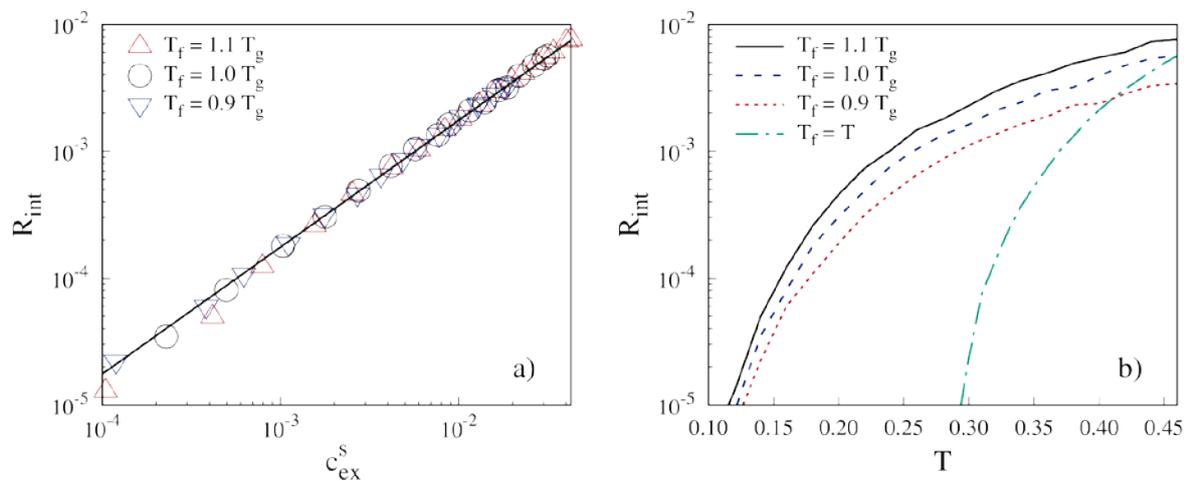

**Figure 7.** a) The dependence of $R_{int}$ on $c_{ex}^s$ for non-equilibrium glasses characterized by fictive temperatures $T_f = 0.9T_g$, $T_g$ and $1.1T_g$. The straight line corresponds to $R_{int} = 0.18\, c_{ex}^s$. b) The dependence of $R_{int}$ on T for the different non-equilibrium glasses. Included, for comparison, is $R_{int}$ for the equilibrium glass (i.e. $T_f = T$).

In summary, we have shown that the intrinsic diffusion rate $R_{int}$ is proportional to the concentration of solute particles on the surface of the glass that are kinetically allowed to exchange with the solvent. This concentration vanishes at a non-zero temperature as a consequence of the approach to a broken ergodicity, a transition that exhibits a weak size dependence. As discussed in the next Section, glasses lack the extended defects like screw dislocations in crystals that dominate dissolution kinetics and so that the disappearance of exchangeable particles means the end all possible routes to dissolution. Changing the fictive temperature alters the dissolution rate by altering the concentration of exchangeable solute particles.

A remaining question, already alluded to, is the relationship between the intrinsic dissolution rate and the actual rate as measured by experiments such as that associated with Fig. 2. The observed dissolution rate is the result of the intrinsic dissolution process, the reverse process of precipitation from solution and the diffusive transport of solute away from the interface. The Stefan problem[29-31] describes the propagation of an interface coupled to a diffusive field where the interface is represented as sharp boundary and its properties reduced to a condition on the solute concentration and its gradient at the boundary. It is not clear whether the details of the intrinsic dissolution of the amorphous solid reported here could be satisfactorily captured by a boundary condition on the solute flux alone. Here we consider a purely numerical approach to the question. Consider imposing a planar boundary within the solution, parallel to the amorphous interface and separated from it by a distance $d$, at which the solute concentration is set to zero. This is achieved by transforming any solute particle that passes through this boundary into a solvent particle. This is not the same criterion that we have applied so far to define the intrinsic rate so far but it does provide, through the distance $d$, a means of continuously varying the nature of the constraint. In Fig. 8 we plot the dissolution rate as a function of $d$, the distance between the amorphous surfaces and the c = 0 boundary. The dissolution rate has been calculated as the steady state rate of solute particles removed per unit surface area, similarly to $R_{int}$.

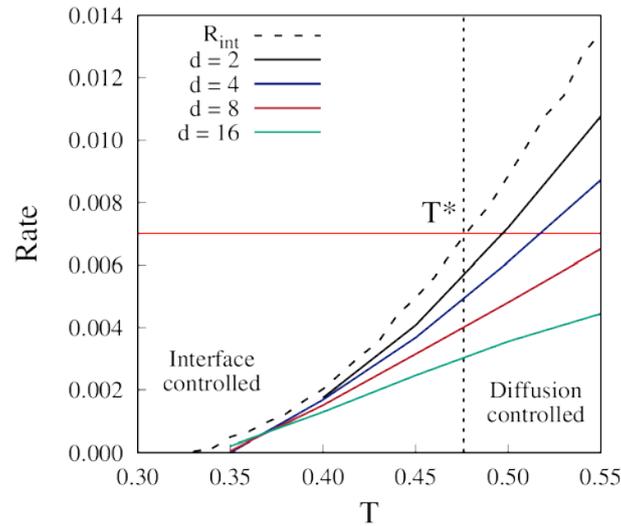

**Figure 8.** The dissolution rate vs temperature for different choices of the distance d (as indicated) between the interface and the absorbing boundary. The intrinsic rate $R_{int}$ is indicated by the black dashed line. The horizontal line indicates an estimate of the high T asymptotic dissolution rate at large $d$ when the dissolution rate is determined by diffusion of solutes from the surface. The vertical line indicates the crossover temperature $T^*$, marking the crossover between interface controlled and diffusion controlled dissolution kinetics.

The dissolution rate is determined by the slowest processes at that temperature. For large d and high T, the dissolution rate is determined by the diffusion of solute from the initial solid region. In the present model, this process is essentially independent of T. Experimentally, we would expect this process to exhibit a much weaker temperature dependence that interface controlled process with its sensitivity to the concentration $c_{ex}^s$. We have estimated this rate from the high T asymptote of the d = 16 simulation and indicated the rate as a horizontal line in Fig. 8. The temperature $T^*$ where this horizontal line intersects the intrinsic dissolution rate $R_{int}$ represents the crossover from interface controlled dissolution at low T to diffusion controlled dissolution at high T (see Fig. 8). Using this construction we find a cross over temperature of $T^* \sim 0.47$ or $\sim 1.03 T_g$, suggesting that the intrinsic dissolution rate provides a reasonable measure of dissolution kinetics over the physically meaningful temperature range (i.e $T < T_g$) when dealing with amorphous solids.

## 5. Morphology of the Solvent-Etched Interface of an Amorphous Solid

In crystal dissolution, the presence of crystal defects in the surfaces provide sites of enhanced dissolution rates that result in the formation of pits[38-41]. Defects at a surface can be characterised by two length scales: their average separation (i.e. defect density) and the average persistence length of a defect into the bulk (i.e the defect's extent). This latter length determines the depth of the associated etch pit. In the case of a screw dislocation in a crystal, this length and, hence, the associated pit depth, can be very large indeed[42,43]. In amorphous materials, dynamic heterogeneities represent a generalization of the idea of a defect, here identified, not be a specific structural feature, but by the local kinetic enhancement. Our goal in this Section is to characterise the surface morphology of the amorphous solid during dissolution and to explore the connection between these surface structures and the underlying inhomogeneities in kinetics that are responsible.

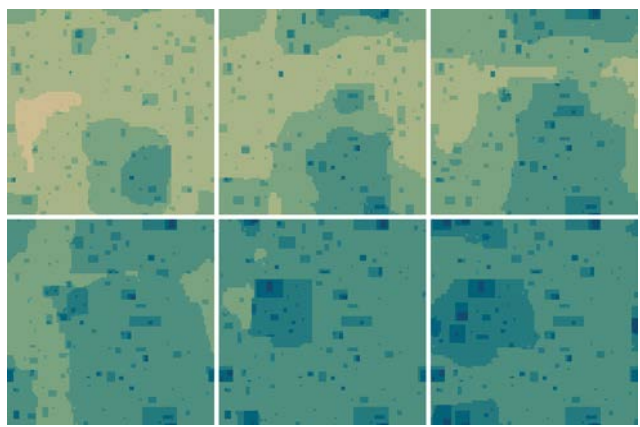

**Figure 9.** The evolution of the surface of the amorphous solute during dissolution at $T = T_f = 0.33$ with each frame separated by a time interval of 2000 MC cycles. The time sequence moves from left to right from the upper left to lower right frames. Solute particles in a given layer are depicted by a colour – light green corresponding the upper most layer and dark blue

the lowest. Note that individual 'pits' do not grow significantly in size as the removal of a couple of layers is sufficient to expose new dissolution sites.

An example of the evolution morphology of the amorphous surface undergoing dissolution is shown in Fig. 9. The roughening of the surface is modest, extending here only over 4 layers. We find no evidence of the fluctuation in height growing significantly with time. This is a consequence of the small spatial extent of individual dynamic heterogeneities. Once dissolved there is nothing left to continue the dissolution preferentially at a given site. Exchange of solvent for solute is found to occur at the flippable sites, by construction, but we observe that only a subset of those solvent domains go on to grow with time. This is analogous to the patterns of relaxation in the bulk of the fkI model[44]. The morphology of our amorphous surface during dissolution is best characterised as a sequential removal of glassy 'layers', initiated around regions of high mobility and then propagated laterally at rates larger than the rate at which the solute-solvent exchange propagates normally into the glass. The dominance of the lateral propagation of dissolution serves to erase the information about the underlying dynamic heterogeneities contained in the location of the initial dissolution sites.

Given this rapid erasure of the initial sites of solvent exchange during dissolution, is there any prospect of using dissolution morphology to directly image the underlying dynamic heterogeneities? The demonstration in the previous Section that the dissolution rate $R_{int}$ vanishes at some characteristic temperature offers one possible approach since, at this temperature, the only solvent exchange possible is that associated with the inherent extractable solute particles. In Fig. 10, we compare the surface distribution of mobile sites and that of solvent penetration. We find that the solvent penetrants have marked out most of the main dynamic domains and provide an excellent direct characterisation of the length scales of the intrinsic dynamic heterogeneities. We emphasise that the while the left panel of Fig. 10 is a dynamic distribution with little chance of direct observation, the right panel is the solvent etched surface that can, in principle, be observed using surface force microscopy. Whether such features would be resolvable on interfaces with thermal roughening is not clear but the possibility that surface force measurements at $T_o$ might provide direct information of dynamic heterogeneities is certainly worth exploring

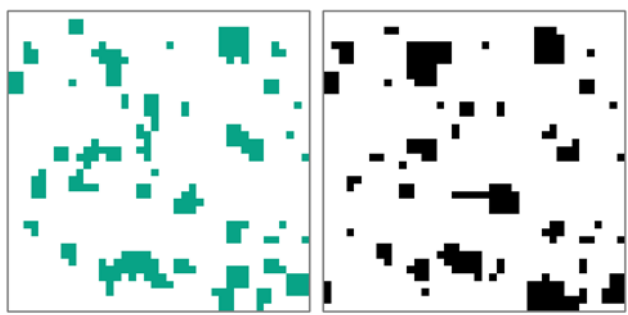

**Figure 10.** The direct visualization of dynamic heterogeneities through solvent etching of the amorphous solid surface. A sample is prepared with a $T_f = 0.414 = 0.9T_g$ and then cooled to $T = 0.11$. The left panel shows the flippable sites (teal) after flipping all initial flippable sites once while the right panel shows the solvent penetration (black) after $10^5$ iterations.

## 6. Conclusions

In this study we have modelled the dissolution process, i.e. the kinetics of the extraction of the solute from the solid into solution, as being controlled by the same type of cooperative processes that control the kinetics of structural relaxation in the amorphous solid. The results of this model are as follows:

i) The intrinsic dissolution rate is proportional to the concentration of surface particles that are kinetically available for exchange with the solvent. This result irrespective of whether the solid is at equilibrium (T = $T_f$) or out of equilibrium ($T \neq T_f$).

ii) When the surface area is finite, the concentration of kinetically removable solute particles is significantly less than that predicted from the bulk concentration (i.e. Eq. 9) at low T due to the arrest (i.e. non-ergodicity) of the surface resulting from the dissolution of these particles. The critical concentration of spins that marks this crossover to non-ergodicity will decrease with the system dimension L as 1/L.

iii) There exists a cross over temperature ($T^*$ from Fig. 8) below which the dissolution kinetics is dominated by the small intrinsic dissolution rate. Above $T^*$, the growth rate is dominated by the diffusion of the solute away from the surface into the solution. In the model studied $T^*$ is found to be ~ $T_g$, indicating that the intrinsic dissolution rate dominates in the temperature regime of interest (i.e. T < $T_g$) where the pure amorphous solute behaves as a solid.

iv) In general, we see expect no macroscopic dissolution pits as are observed in crystal dissolution. The kinetic heterogeneities of the amorphous solid are small in dimension and their position is quickly erased by the removal of solute into the solution.

In practical terms, if one's goal is to increase the solubility rate of an amorphous solid our results indicate that this must be accomplished by increasing $c_{ex}^s$ and, hence, the fictive temperature of the glass. Rapid quenching of a melt or mechanical grinding of the solid are two means to this end with a likely increase in dissolution rates, as we have shown, of $\leq 2$.

The selective etching of solutes from local sites of kinetic facilitation raises the intriguing possibility of measuring the topology of the dissolving surface to determine the size and density of these otherwise invisible objects. As noted above, these objects are typically small in size and so their presence is erased with the removal of a few layers of solute. As we have shown, however, information about the inherent heterogeneities is still accessible in the form of the small etch pits visible in surfaces studied near or below the critical temperature $T_o$ where the dissolution process is sufficiently slow.

In this first study, we have restricted ourselves to a pure glass-forming material. When the glass consists of multiple components with different solubilities, dissolution will be dominated by the selective removal of the most soluble species. Selective dissolution can dramatically change the properties of the glass surface, producing a gel, in the case of selective ion removal from silicate glasses[6-9], or a nanoporous surface, in the case of electrochemical or solvent-selective dissolution from a metallic glass[4,5]. The combination of compositional organization and dynamic heterogeneity that must contribute to the observed surface changes highlight the rich range of chemical phenomena (and the associated physical information) accessible through detailed study of the dissolution of an amorphous solid.

**Acknowledgements**

We acknowledge funding support for this research from the Australian Research Council and the provision of computer time from the High Performance Computing Facility at the University of Sydney. One of us (PH) gratefully acknowledges Mark Ediger for bringing the phenomenon of Class II diffusion to our attention and so providing the initial motivation for this study.

**Appendix: Dynamic Asymmetry and The Crossover from Dissolution to Inter-Diffusion**

In studying glass dissolution, we have been restricted, by definition, to $T < T_g$. We can, however, still ask how does the process of dissolution crossover, as T is increased above $T_g$, into the processes of liquid-liquid inter-diffusion? The glass transition itself corresponds to the continuous transition between two states, liquid and solid, that are typically regarded as non-overlapping. The transition between dissolution and inter-diffusion captures an analogous dissonance.

The essential distinction between dissolution and inter-diffusion lies in the asymmetry of diffusion rates of the minority species in the solution and solid phases. In solid dissolution, the solid is essentially impermeable to the solvent so that the mixing occurs entirely within the solution phase. The extraction of the solute from the solid is, as described in the previous Sections, a complex process, sensitively dependent on the kinetic fluctuations at the glass surface. As we increase the temperature beyond $T_g$ we observe an increasing rate of penetration of the solvent into the solid and an associated reduction in the influence of the dynamic heterogeneities of the solute-rich phase.

To model this crossover we shall return to the original model in which the concentration of the solute in solution is not constrained. This means that dissolution kinetics is coupled to diffusion in the solvent and, as we increase T, in the solute rich phase as well. In our model, the self diffusion coefficient $D_s$ of the solute in the dilute solution is constant, independent of T. In contrast, the self diffusion coefficient $D_g$ of a single solvent particle in the glass/solute-rich phase is highly T dependent. In Fig. 11, the asymmetry of dynamics in the two phases can be characterised by the ratio $D_g/D_s$. We find that $\dfrac{D_g}{D_s} \propto \exp(-T_{asym}/T)$ with $T_{asym} \sim 8$.

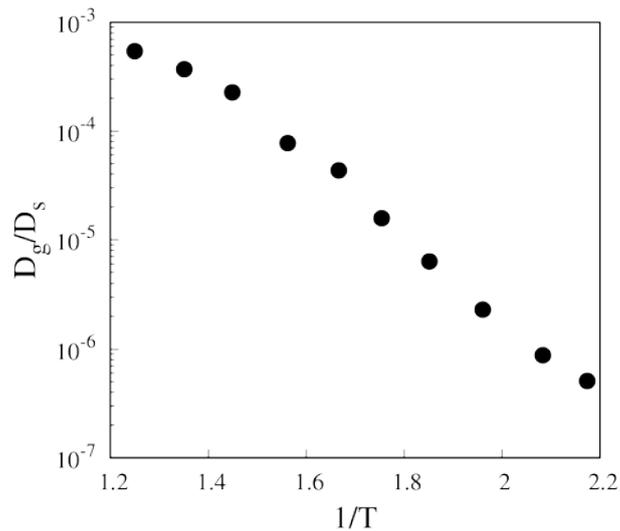

**Figure 11.** A plot of the dynamic asymmetry $D_g/D_s$ vs $1/T$ as a log-linear plot.

The dynamic asymmetry between the solvent and solute phases and its temperature dependence, shown in Fig. 11, captures the essential crossover in the time evolution of the concentration gradient of the interface between the two phases in going between liquid and glassy solute. In the case of symmetric dynamics we have diffusive mixing with a maximum gradient that decreases monotonically with time. Dynamic asymmetry, on the other hand, results in solid dissolution and an interface that propagates as a narrow front, characterised by a maximum gradient that shows little change as the mixing proceeds. In Fig. 12 we plot the concentration profiles for a sequence of times for three different temperatures, each associated with a different value of the asymmetry ratio $D_g/D_s$.

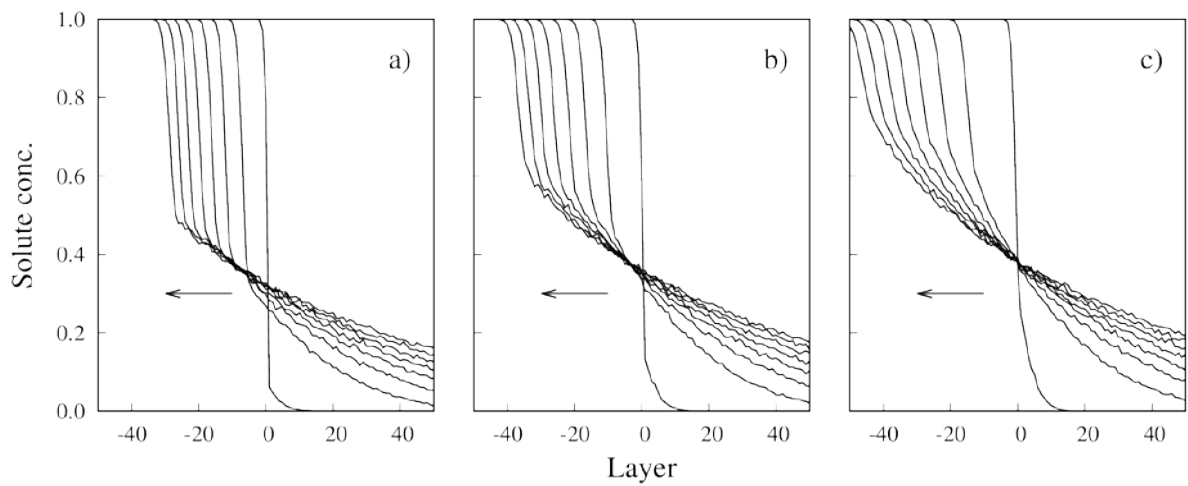

**Figure 12.** Plots the time evolution of the solute concentration for three different dynamic asymmetries i.e. $D_g/D_s = 9 \times 10^{-7}$ (a), $1.6 \times 10^{-5}$ (b) and $2.3 \times 10^{-4}$ (c). Note the transition from front-like propagation in the left panel to diffusive mixing in the right panel. The arrows indicate the direction of front propagation.

Between the two well defined limits, the dissolution of an impermeable solid and the diffusive mixing of two similar liquids, we can find situations of intermediate asymmetry where some solvent penetration of the amorphous solid occurs.

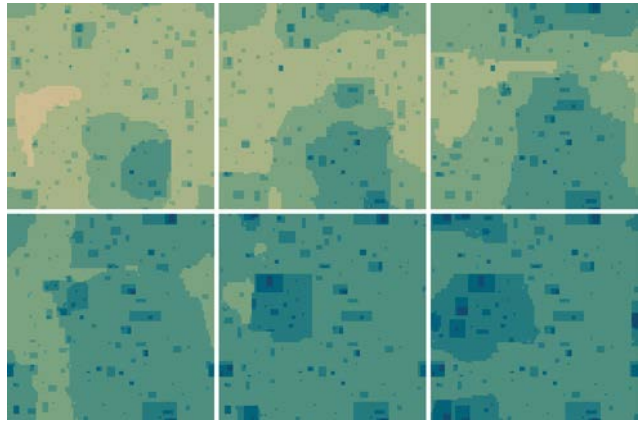

TOC graphic